\documentclass[useAMS,usenatbib]{mn2e}

\newcommand{\be} {\begin{equation}}

\newcommand{\ee} {\end{equation}}

\newcommand{\isaac}{{\sl ISAAC}}
\newcommand{\vlt}{{\sl VLT}}
\newcommand{\fors}{{\sl FORS1}}
\newcommand{\twomass}{{\sl 2MASS}}
\newcommand{\cxo}{{\it Chandra}}

\newcommand{\rbs}{RBS\,1774}
\newcommand{\CXO}{{\it Chandra}}
\newcommand{\R}{{\it ROSAT}}

\newcommand{\xmm}{{\it XMM--Newton}}

\newcommand{\bc}{\begin{center}}
\newcommand{\ec}{\end{center}}
\newcommand{\ltsima}{\; \buildrel < \over \sim \;}
\newcommand{\lsim}{\lower.5ex\hbox{\ltsima}}
\newcommand{\loe}{\lower.5ex\hbox{\ltsima}}
\newcommand{\gtsima}{\; \buildrel > \over \sim \;}
\newcommand{\gsim}{\lower.5ex\hbox{\gtsima}}
\newcommand{\goe}{\lower.5ex\hbox{\gtsima}}

\def\ergscm2 {erg\,s$^{-1}$cm$^{-2}$}

\def\farcs{\hbox{$.\!\!^{\prime\prime}$}}

\def \oneight {\hbox{RX\, J1856.5-3754}}
\def \zeroseven {\hbox{RX\, J0720.4-3125}}
\def \zerofour {\hbox{RX\, J0420.0-5022}}
\def \zeroeight {\hbox{RX\, J0806.4-4123}}
\def \onethree{\hbox{RX\, J1308.6+2127}}
\def \onesix{\hbox{RX\, J1605.3+3249}}
\def \seventeen{\hbox{1RXS\, J214303.7+065419}}

\title[X-ray position and multiwavelength observations of RBS\,1774]{Accurate X-ray position and multiwavelength observations of the isolated neutron star RBS\,1774}

\author[Rea et al.]{N. Rea$^{1,2}$\thanks{E-mail: N.Rea@sron.nl}, 
M.A.P. Torres$^{3}$, P.G. Jonker$^{1,3,4}$, R.P. Mignani$^{5}$, S. Zane$^{5}$, M. Burgay$^{6}$, \newauthor D.L. Kaplan$^{7}$, R. Turolla$^{8,5}$, G.L. Israel$^{9}$, D. Steeghs$^{3}$ \\
$^{1}$SRON Netherlands Institute for Space Research, Sorbonnelaan, 2, 3584CA, Utrecht, The Netherlands \\
$^{2}$School of Physics, A29, University of Sydney, NSW 2006, Australia \\
$^{3}$Harvard-Smithsonian Center for Astrophysics, 60 Garden Street, Cambridge, MA 02138, USA \\
$^{4}$Astronomical Institute, Utrecht University, PO Box 80000, 3508 TA, Utrecht, The Netherlands \\
$^{5}$Mullard Space Science Laboratory, University College London, Holmbury St. Mary, Dorking Surrey, RH5 6NT, UK \\
$^{6}$INAF--Astronomical Observatory of Cagliari, Loc. Poggio dei Pini, Strada 54, 09012 Capoterra (CA), Italy \\
$^{7}$ Kavli Institute for Astrophysics and Space Research, 
Massachussets Institute of Technology, Cambridge, MA 02139, USA \\
$^{8}$ Department of Physics, University of Padua, via Marzolo 8, I-35131, Padova, Italy \\ 
$^{9}$INAF--Astronomical Observatory of Rome, via Frascati 33, 00040, 
Monteporzio Catone (RM), Italy }

\input psfig.sty

\begin{document}

\pagerange{\pageref{firstpage}--\pageref{lastpage}} \pubyear{2006}

\maketitle

\label{firstpage}

\begin{abstract}

  We report on X--ray, optical, infrared and radio observations of the
  X-ray dim isolated neutron star (XDINS) 1RXS J214303.7+065419 (also
  known as \rbs). The X--ray observation was performed with the High
  Resolution Camera on board of the \CXO\, X-ray Observatory, allowing
  us to derive the most accurate position for this source ($\alpha$ =
  21$^{\rm h}$43$^{\rm m}$3$^{\rm s}$.38, $\delta$ =
  +6$^{\circ}$54$^{\prime}$17$\farcs$53; 90\% uncertainty of
  $0\farcs6$). Furthermore, we confirmed with a higher spatial
  accuracy the point--like nature of this X--ray source. Optical and
  infrared observations were taken in B, V, $r^{\prime}$,
  $i^{\prime}$, J, H and K$_s$ filters using the Keck, VLT, Blanco and
  Magellan telescopes, while radio observations were obtained from the
  ATNF Parkes single dish at 2.9\,GHz and 708\,MHz. No plausible
  optical and/or infrared counterpart for \rbs\, was detected within
  the refined sub--arsecond \CXO\, X--ray error circle. Present upper
  limits to the optical and infrared magnitudes are $r^{\prime}>$25.7
  and J$>$22.6 (5$\sigma$ confidence level). Radio observations did
  not show evidence for radio pulsations down to a luminosity at
  1.4\,GHz of $L < 0.02$\,mJy\,kpc$^2$, the deepest limit up to date
  for any XDINS, and lower than what expected for the majority of
  radio pulsars. We can hence conclude that, if \rbs\, is active as
  radio pulsar, its non detection is more probably due to a
  geometrical bias rather than to a luminosity bias. Furthermore, no
  convincing evidence for RRAT--like radio bursts have been found. Our
  results on \rbs\, are discussed and compared with the known
  properties of other thermally emitting neutron stars and of the
  radio pulsar population.

\end{abstract}


\section{Introduction}

Over the last decade, thank to the high sensitivity of \R\, in the
soft X--ray band (0.1--2\,keV), a number of thermally emitting X-ray
pulsars have been discovered, commonly called X--ray dim isolated
neutron star (XDINSs). To date, seven such sources are known:
\oneight,
\zeroseven, \zerofour, \zeroeight, \onethree, \onesix\, and
\seventeen\, (see Haberl~2007 and van~Kerkwijk \& Kaplan~2007 for
recent reviews).

%%%%%%%%%%%%%%%%%%%%%%%%%%%%%%%%%%%%%%%%%%%%%%%%%%%%%%%%%%%%%%%%%%%%%%%%%%%%%%

\begin{table*}
\begin{center}
\begin{tabular}{cccccc}

\hline
\hline 
  \multicolumn{1}{c}{\underline{Instrument}} & \multicolumn{1}{c}{\underline{Date (UT)}} & \multicolumn{1}{c}{\underline{Exposure (ks)}} & \multicolumn{1}{c}{\underline{Band}} &  \multicolumn{1}{c}{\underline{Pixel Size ($^{\prime\prime}$)}} \\

 &  & & &  \\
\CXO\,HRC--I & 2006-07-20  & 9.7 & 0.3--8\,keV & 0.131 \\

 &  & & &  \\
& &  & \multicolumn{1}{c}{\underline{Filter}} &  \multicolumn{1}{c}{\underline{Seeing ($^{\prime\prime}$)}} \\   
&  & & &  \\
Keck    & 2001-09-21 & 2.06 & B  & 0.5 \\
Blanco  & 2006-06-29  & 2.4 & r$^{\prime}$ & 1.1--1.2  \\
        & 2006-06-29  & 1.44 & i$^{\prime}$ & 1.1--1.2  \\
VLT & 2005-08-31 & 3.0 & V & 1.5 \\ 
     & 2003-11/2004-01 & 2.6  & H  & 0.6--0.9 \\

Magellan & 2006-08-6,10,13 & 9.0 & J   & 0.5 \\
         & 2006-08-6,7,8 & 9.72 & K$_s$   & 1 \\
&  & & & \\

 &  & & \multicolumn{1}{c}{\underline{$\nu$ (MHz)}}   & \multicolumn{1}{c}{\underline{$\Delta\nu$ \& $\delta\nu$ (MHz)}}   \\
&  & & &  \\
Parkes  & 2006-04-20  & 12.96 & 2935.5  & 576.0 , 3.0   \\
    &    2006-04-20  & 12.96  & 708.875 &  64.0 , 0.25 \\          
\hline
\hline 
\end{tabular}

\caption{Log of the observations.}
\label{obstable}
\end{center}
\end{table*}

%%%%%%%%%%%%%%%%%%%%%%%%%%%%%%%%%%%%%%%%%%%%%%%%%%%%%%%%%%%%%%%%%%%%%%%%%%%%%%

There is now consensus that XDINSs are nearby, middle-aged ($\approx
10^6$~yr), cooling neutron stars. Their properties are, however, at
variance with those of radio pulsars, and with those of other classes
of isolated neutron stars detected at X-ray energies.  In particular,
i) their spectra are purely thermal, with no evidence for a power-law
tail extending to higher energies; ii) their spin periods cluster in a
rather restricted range ($\sim 3$--11\,s) and are much longer than
those typical of radio pulsars (similar, however, to those of the
magnetar candidates, namely the Anomalous X-ray Pulsars (AXPs) and the
Soft Gamma-Repeaters (SGRs), e.g. Woods \& Thompson~2006); and iii) no
evidence for radio emission has been reported so far despite deep
searches (Brazier \& Johnson~1999; Johnston~2003; Kaplan et al.~2003;
Burgay et al.~2007, in prep). Recently, pulsed emission from two
sources have been claimed at very low frequencies (Malofeev et
al.~2005, 2007), but this has not been confirmed yet .

The low values of the column density derived from X-ray data
($N_H\approx 10^{20}\ {\rm cm}^{-2}$) indicate that XDINSs have a
typical distance of a few hundred parsecs (see e.g. Posselt et
al.~2007). The overall X-ray spectrum of XDINS is remarkably well
reproduced by an absorbed blackbody with temperatures in the range
$kT\sim 40$--100\,eV. Application of more sophisticated, and
physically motivated, models for the surface emission (atmospheric
models in particular) result in worse agreement with the
data. Furthermore, \xmm\, and \CXO\, observations have shown that in
all XDINS spectra, except \oneight, broad ($EW\approx 10$--100~eV)
absorption features are present at energies of several hundred
eVs. The most likely interpretations are that they are due to proton
cyclotron (at the fundamental resonance) and/or bound-free,
bound-bound transitions in H, H-like and He-like atoms in the presence
of a relatively high magnetic field $B\approx 10^{13}$--$10^{14}$~G.

%For such large field strengths, exotic molecular systems and broad
%spectral edges originating from the solid crust are also expected
%(Turbiner et al.~2007; Turbiner \& Lopez-Vieyra 2006; P\'erez-Aror\'in
%et al.~2006).  The magnetic field strengths inferred from the
%absorption features (a few $10^{13}$~G) are in reasonable agreement
%with those determined from the spin-down of the only two XDINSs having
%up to now an X--ray timing solution: \zeroseven\, and
%\onethree\, (Kaplan \& van Kerkwijk 2005a, b).

The detection of an optical counterpart is fundamental in XDINS
science, because it paves the way to the measure of the neutron star
proper motion (as in RX J1856.5-3754, Walter 2001, Neuh\"auser 2001;
RX J0720.4-3125, Motch et al. 2003; RX J1605.3+3249, Motch et
al. 2005, Zane et al.~2006), and of the parallax, as in RX
J1856.5-3754 (Walter \& Lattimer 2002, van Kerkwijk \& Kaplan 2007)
and RX J0720.4-3125 (Kaplan et al.~2007). The knowledge of the
distance, coupled with X-ray data, may provide tight constraints on
the star radius allowing to test the equation of state of the matter
at supra-nuclear densities.  Furthermore, because of the emission
coming directly from the star surface, XDINSs offer an unprecedented
opportunity to investigate the thermal and magnetic distributions of
isolated neutron stars. Up to now, however, no self-consistent model
can properly account for the multiwavelength spectral energy
distribution (SED) of XDINSs, despite some recent progresses (e.g. Pons
2002; Turolla et al. 2004; Geppert et al. 2006; Zane \& Turolla 2006;
P\`{e}rez-Azorin al. 2006; Ho et al. 2007).

%Despite some progresses (e.g. Pons 2002; Turolla et al. 2004; Geppert
%et al. 2006; Zane \& Turolla 2006; P\`{e}rez-Azorin al. 2006), there
%are several issues that still need further investigation: i.e. how to
%produce a purely blackbody spectrum at X-ray energies, the origin of
%spectral features and of the optical excess. In this respect
%multiwalength observations of these nearby hot neutron stars are of
%paramount importance not only to shed light on the nature of this
%group of isolated neutron stars, but also to understand the neutron
%star class in general.

1RXS~J214303.7+065419 (\rbs) was identified in the \R\, Bright Source
catalogue at about 48$^{\prime}$ from the BL Lac MSS~2143.4+0704
(Zampieri et al.~2001). Despite the limited statistics, this first
\R\, PSPC observation was enough to reveal the thermal character of
the X-ray spectrum ($kT\sim 92$~eV). This, together with a lower limit
of $\sim 1000$ obtained for the optical-to-X-ray flux ratio, fully
qualified \rbs\, as a XDINS candidate. A subsequent \xmm\, pointed
observation confirmed \rbs\, as a member of the XDINS class and
revealed a spin period of $\sim$9.437\,s with a pulsed fraction of
$\sim$4\% (Zane et al.~2005). Similar to other XDINS, the \xmm\,
spectrum of \rbs\, is well fitted by an absorbed blackbody ($N_H\sim
3.65\times10^{20}$\,cm$^{-2}$, $kT\sim104 $\,eV) plus an absorption
edge at $E_{edge}\sim 694$\,eV (see Zane et al.~2005, Cropper et
al.~2007 for further details). The 0.2--2\,kev unabsorbed flux was
$\sim5\times10^{-12}$\ergscm2 .

Optical follow up with the New Technology Telescope (NTT) in La Silla
(Chile), revealed many sources in the \R\, error circle (see tab.\,2
in Zampieri et al.~2001) up to a limiting magnitude of
R$\sim$22.8. However, the large \R\, \rbs\, position uncertainty
prevented to reliably accept or reject any candidate, except the
brightest ones, the color of which were inconsistent with being the
counterpart of an isolated neutron star. In the smaller \xmm\,
error circle no optical counterpart was found in the NTT exposure
(Zane et al.~2005). More recently, Mignani et al.~(2007) reported on
\vlt\ observations that revealed no source in the \xmm\, error
box, down to a limiting magnitude of $V \sim 25.5$.

We report in \S\,\ref{chandra} on the accurate X-ray position of
\rbs\, obtained with the Imaging detector of the High Resolution
Camera (HRC-I) on board of the \CXO\, X-ray Observatory. In
\S\,\ref{keck} we show optical and infrared observations of the field
around \rbs, and in \S\,\ref{parkes} we present deep radio
observations taken with the ATNF Parkes single dish
antenna\footnote{Through this paper we assumed for \rbs\, a distance
of 400\,pc, based on $N_H$ measurements from Posselt et
al.~(2007). However, our results will need to be scaled if a more
exact measurement of the distance will become available.}. We then
present our results in \ref{results}, and discuss our findings in
\S\,\ref{discussion} in comparison with those observed in other
thermally emitting neutron stars.

%%%%%%%%%%%%%%%%%%%%% FIGURE PSF %%%%%%%%%%%%%%%%%%%%%%%%%55
\begin{figure}
\centerline{\psfig{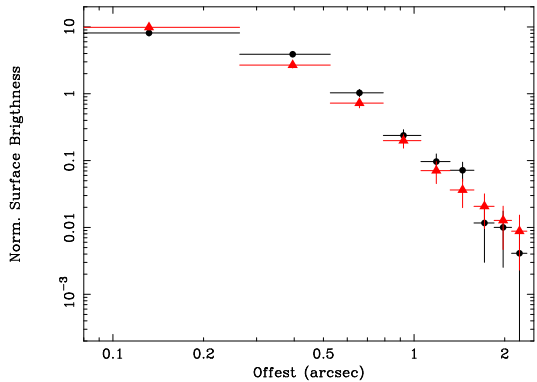}}
\caption{\CXO\, HRC--I \rbs\, 2D profile (filled circles) superimposed with the simulated instrumental PSF (filled triangles).}
\label{profile}
\end{figure}

%%%%%%%%%%%%%%%%%%%%%%%%%%%%%%%%%%%%%%%%%%%%%%%%%%%%%%%%%%%%%%%%

\section{Observations}

\subsection{X-ray observation: {\it Chandra}}
\label{chandra}

The \CXO\, High Resolution Imaging Camera (HRC--I; Zombeck et
al. 1995) observed the XDINS \rbs\, on 2006 November 26th, for an
on--source exposure time of $\sim$1.1\,ks. The target was detected
$0^{\prime}.32$ off--axis with respect to the standard aim-point. Data
were reduced with {\tt CIAO 3.4}
software\footnote{http://asc.harvard.edu/ciao/} and analysed with
standard software packages for X--ray data ({\tt Ximage} and {\tt
Xronos}).  We first checked the data for the presence of solar flares
and extracted a new observation--specific bad--pixel file. We then run
a degap correction, cleaned the image for the hot pixels, and
corrected the astrometry for any processing offset.

Only one source was significantly detected in the whole
$30^{\prime}\times30^{\prime}$ HRC--I field of view. We used three
detection algorithms: the CIAO {\tt wavedetect} and {\tt celldetect}
tools, and the wavelet-based PWDETECT software developed and optimised
for \CXO\, images (Damiani et al. 1997). The source has the following
coordinates: $\alpha$ (J2000)= 21$^{\rm h}$43$^{\rm m}$3$^{\rm s}$.38,
$\delta$ (J2000) = +6$^{\circ}$54$^{\prime}$17$\farcs$53. The
statistical uncertainty in the position was 0\farcs18. Unfortunately
the lack of other sources in the field not allow us to correct for
uncertainties in the bore-sight. We therefore took the \CXO\, aspect
solution for HRC--I which gives an uncertainty of 0\farcs6 (90\%
confidence level). The \rbs\, position was consistent within all the
detection algorithms and with the previous \R\, and \xmm\, error
circles (Zampieri et al.~2001; Zane et al.~2005). Finally, the spatial
profile was found to be in good agreement with the expected \CXO\,
point spread function (PSF) for on--axis source (see
Fig.\,\ref{profile}).

Photon arrival times were extracted from a circular region with a
radius of 3$^{\prime\prime}$, including more than 90\% of the source
photons, and corrected to the barycenter of the Solar System. We
collected a total of 325 counts from the source, we then inferred an
effective HRC--I countrate of $0.290\pm0.017$\,ct\,s$^{-1}$. Assuming
that the \rbs\, continuum spectrum did not change with respect to the
latest \xmm\, observation ($N_H\sim3.65\times10^{20}$\,cm$^{-2}$ and
kT$\sim104$\,eV; Zane et al. 2005), we inferred (using {\tt
WebPIMMS}\footnote{http://heasarc.gsfc.nasa.gov/cgi-bin/Tools/w3pimms})
an absorbed 0.2--2\,keV flux of $2.6\times10^{-12}$\,\ergscm2 , which
translates to an unabsorbed flux of $4.7\times10^{-12}$\,\ergscm2
. This flux is consistent with the source having remained stable.

Despite the very accurate timing resolution of the HRC--I camera, no
pulsations at the period of $\sim$9.4\,s (or any other period) were
detected, due to the low number of counts. In fact, no pulsations at
the predicted spin period would had been detectable unless having a
pulsed fraction $>$80\%. \rbs\, pulsed fraction is 4\%, well below
this limit.

%%%%%%%%%%%%%%%%%%%%%%%% IR field %%%%%%%%%%%%%%%%%%%%%%%%%%
%\begin{figure*}
%\centerline{
%\psfig{figure=finding_chart_blanco_new2_r.ps,width=15cm}}
%\caption{Finding chart of the field of \rbs\, taken in the r$^{\prime}$-band with the Blanco telescope (see Sec.\ref{blanco}). Horizontal black lines are artifacts due to very bright stars, and RASS error circle around the position of \rbs\, (90\% confidence level; Zampieri et al.~2001) is overplotted for clarity.}
%\label{findchart}
%\end{figure*}

%%%%%%%%%%%%%%%%%%%%%%%%%%%%%%%%%%%%%%%%%%%%%%%%%%%%%%%%%%%%%

%21 43 3.3864   6 54 17.532 0.185576 0.27487E+00+/-0.17017E-01 0.30

\subsection{Optical and infrared observations: \\ {\it Keck, VLT, Blanco} and {\it Magellan}}
\label{keck}
\label{blanco}
\label{vlt}
\label{magellan}

\begin{figure*}
\vbox{
\hbox{\centerline{
\psfig{figure=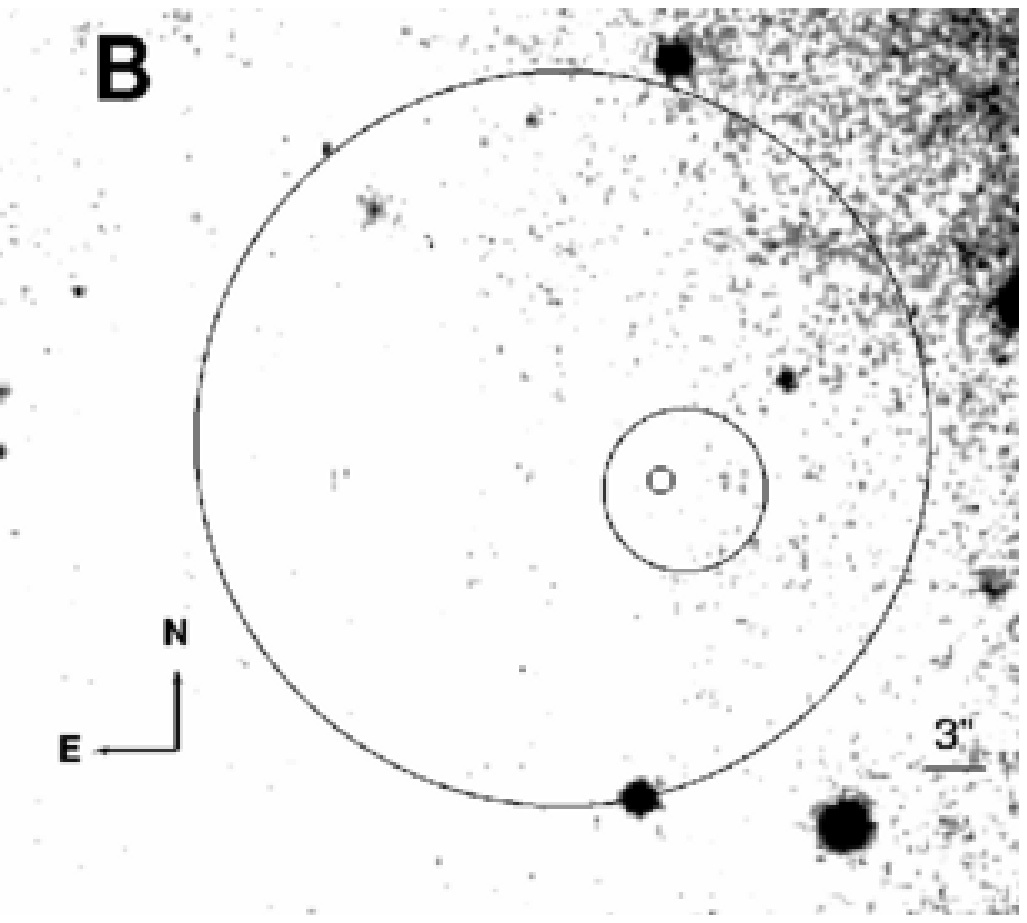,width=6cm}
\hspace{0.3cm}
\psfig{figure=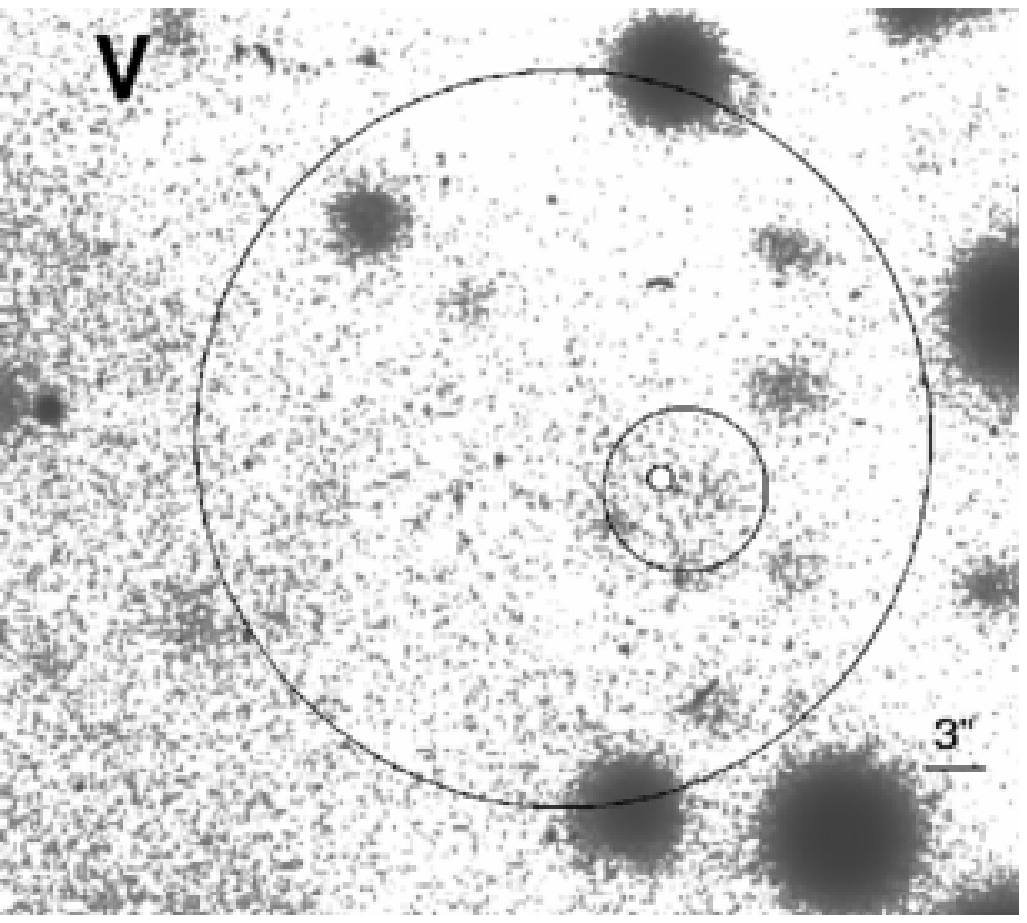,width=6cm}
\hspace{0.3cm}
\psfig{figure=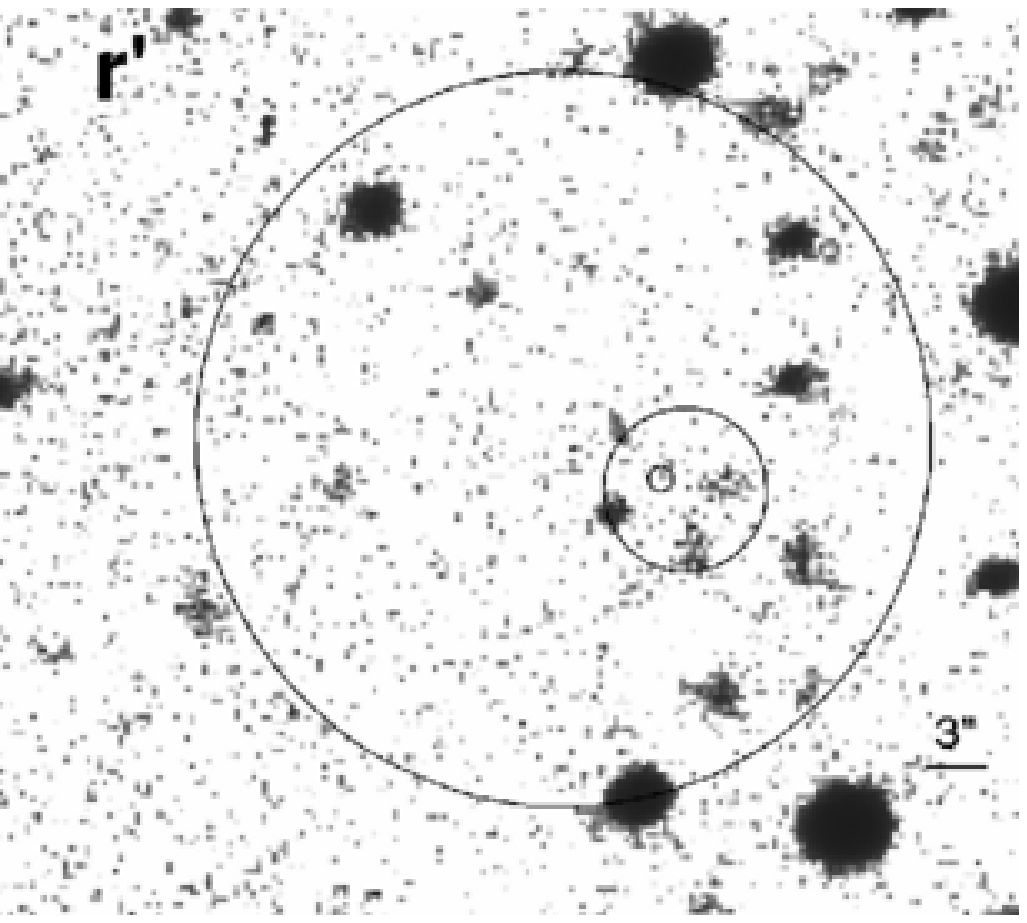,width=6cm} }}
\vspace{0.5cm}
\hbox{\centerline{
\psfig{figure=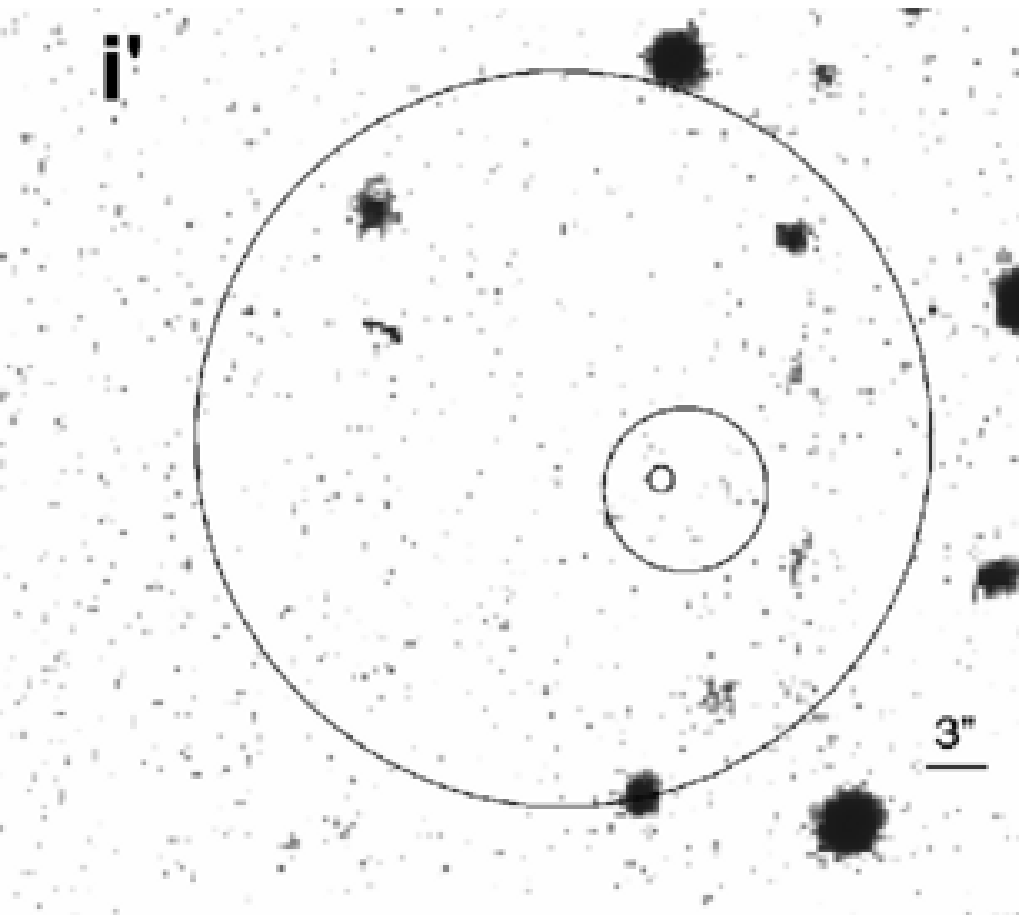,width=6cm} 
\hspace{0.3cm}
%\psfig{figure=magellan_J_low.ps,width=6cm} 
%\hspace{0.3cm}
\psfig{figure=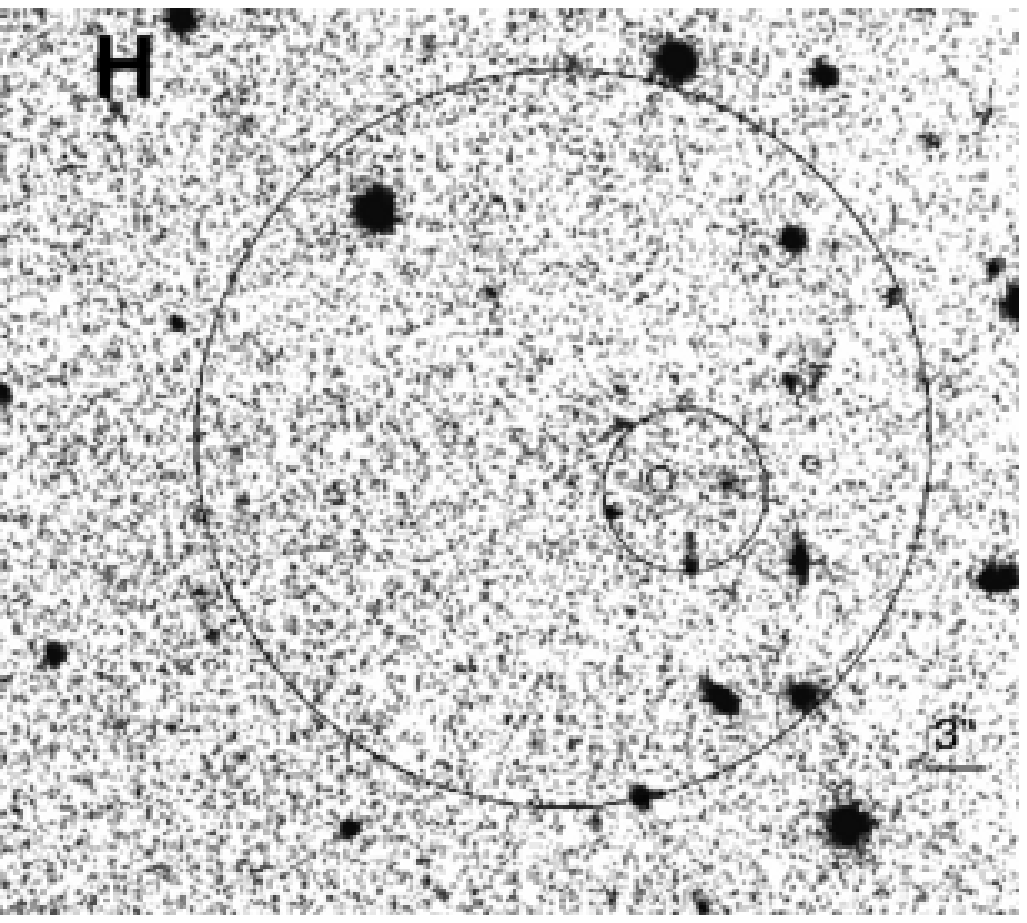,width=6cm}
\hspace{0.3cm}
\psfig{figure=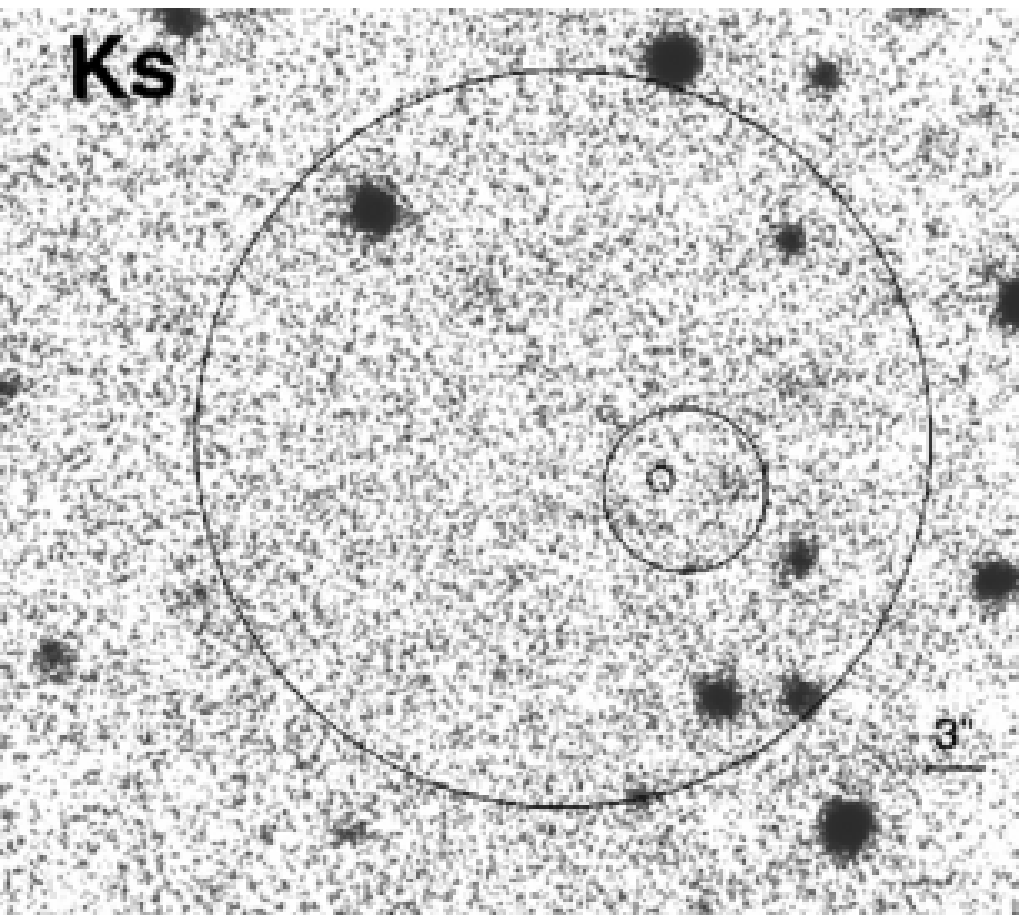,width=6cm}}}}
\caption{Optical and infrared fields of \rbs\, with overplotted the {\it RASS} (Zampieri et al.~2001), \xmm\,(Zane et al.~2005) and \CXO\, 90\% error circles on its position ({\it RASS}$>$\,\xmm\,$>$\,\CXO). We plotted the histogram of the images and smoothed them with a Gaussian function. From left to right, {\em first row}: $B$, $V$ and $r^{\prime}$ images from Keck, VLT and Blanco, respectively; {\em second row}: $i^{\prime}$, $H$ and $K_s$ images from Blanco, VLT and Magellan, respectively. North is up and East left for all images. Note that the few black dots within the \CXO\, error circle, are due to hot pixels.}
\label{iroptrbs}
\end{figure*}

Optical and infrared images of \rbs\, were taken in the $B$, $V$,
$r^\prime$, $i^\prime$, $J$, $H$ and $K_s$ bands (see Tab.\,1 for the
log of the observations and Fig.\,\ref{iroptrbs} for the finding chart).

Images in the $B$--band were obtained on 2001 September 21 using the
LRIS instrument (Oke et al.~1995) mounted on the Keck~I telescope at
the W.M. Keck Observatory in Hawaii. The field was observed in the
B-band for a total exposure of 2060\,s. The night was not photometric
whereas the seeing was 0\farcs5. The data were corrected for bias and
flat field in the usual way using {\it
IRAF}\footnote{http://iraf.noao.edu/}. The photometric zero-point was
determined for the combined frame by measuring 10 stars well-detected
in both the combined frame and the 60-s frame (we could not use the
same zero-point for both due to transparency variations), and checked
by directly measuring 4 USNO B1.0 stars in the field in \rbs\, that
were not saturated in our stacked image. We estimate an uncertainty of
0.5 magnitudes in the zero-point.

On 2006 June 29 we observed the \rbs\, field with the $r^\prime$ and
$i^\prime$ Sloan filters using the MOSAIC--II instrument mounted at
the 4~m Blanco telescope at the Cerro Tololo Inter--American
Observatory. We observed the field for 2400~s in $r^\prime$ and 1440~s
in $i^\prime$ (see also Tab.\,1).  The data were corrected for bias
and flat field in the usual way using the {\it MIDAS}
software\footnote{http://www.eso.org/projects/esomidas/}. The night
was photometric and the seeing was 1\farcs1--1\farcs2. The frames were
calibrated with respect to the standard star SA~107 using the SDSS DR5
magnitudes of that field (Adelman-McCarthy et al. 2007 submitted).
We estimated an error of 0.3 magnitudes in our zero-point.

%%%%%%%%%%%%%%%%%%%%%%%% IR field %%%%%%%%%%%%%%%%%%%%%%%%%%

%%%%%%%%%%%%%%%%%%%%%%%%%%%%%%%%%%%%%%%%%%%%%%%%%%%%%%%%%%%%%

\rbs\, was subsequently observed in the $V$ and $H$-bands from the VLT at 
Paranal Observatory in Chile using \fors\ (FOcal Reducer Spectrograph)
and \isaac\ (Infrared Spectrometer And Array Camera),
respectively. \fors\ was operated in its standard resolution mode
(pixel size $0\farcs2$; field of view $6\farcm8 \times 6\farcm8$).
Out of the original program approved for Service Mode, \rbs\, was
observed in the $V$-band only for about one hour (see also Mignani et
al.~2007), and with relatively bad seeing conditions ($\sim
1\farcs5$). Photometric calibration was ensured by observations of
Landolt stars at the beginning of the night. Data reduction (bias
subtraction, flat fielding) was performed using the \fors\ data
reduction pipeline. The exposures were combined and cosmic rays hits
filtered out. For the \isaac\, observations in the $H$-band, the Short
Wavelength (SW) camera was used, equipped with a Rockwell Hawaii
1024$\times$1024 pixel Hg:Cd:Te array (0\farcs148 pixel size;
152$\times$152 arcsec field of view).  To allow for sky background
subtraction, each exposure was split in sequences of 33 randomly
dithered exposures of 5$\times$12 s each.  Night conditions were good
($0\farcs6-0\farcs9$ seeing) but not perfectly photometric due to the
presence of variable cirrus clouds.  Science exposures were retrieved through
the public ESO archive\footnote{http://archive.eso.org/} together with
the closest in time calibration files and reduced using the ESO's
eclipse
package\footnote{http://www.eso.org/projects/aot/eclipse/eug/eug/eug.html}
for de-jitter and sky subtraction.  Stacks of images taken in
different nights have been co-added.  Due to the lack of suitable
standard star observations during the nights, for the photometric
calibration we have used \twomass\ stars identified in the image. This
yielded a photometric calibration accurate within $\sim0.3$
magnitudes, also accounting for the photometric accuracy of
\twomass\ and the passband difference between the \twomass\ and the
\isaac\ $H$-band filters.

Furthermore, we observed \rbs\, in the $J$ and $K_s$-bands using the
PANIC camera on the 6.5-m Baade Magellan Telescope at Las Campanas
Observatory (LCO). PANIC (the Persson's Auxiliary Nasmyth Infrared
Camera) yields a $0\farcs125$~pixel$^{-1}$ plate scale onto a Rockwell
Hawaii Hg:Cd:Te $1024 \times 1024$ array (Martini et al. 2004). The
observations consisted of several 9 point dither patterns with a 20s
($K_s$) and 60s ($J$) exposure repeated three times at each offset
position. The total time expended on source was 2.5 and 2.7\,h in the
$J$ and $K_s$-bands, respectively. However, weather conditions were
variable (with seeing ranging 0\farcs5 and 1\farcs2) and we had to
reject a significant number of images obtained under bad conditions
(poor seeing, clouds and/or variable background) as including them
would deteriorate the final photometry. The data were reduced through
the PANIC software: the raw frames were first dark subtracted and
flat-fielded. Master flat-fields were made by combining 75 ($J$) and
95 ($K_s$) twilight flat field frames scaled by their mode. Next, a
sky image was subtracted from the set of target frames. The sky image
was built by masking out stars from each set of dithered
frames. Finally, a mosaic image was obtained by combining and
averaging the sky-subtracted images. For the analysis we used mosaic
images generated by stacking the frames with better seeing and not
affected by variable weather. The total on-source time for these
mosaic frames was 54 and 108 min, with a seeing of $0\farcs5$ and
$1^{\prime\prime}$ for $J$ and $K_s$, respectively. Absolute
calibration of the $J$-band data set was performed using the standard
star S301-D (Persson et al. 1998), observed the same night the \rbs\,
$J$ image was obtained, and assuming a median extinction value of
$=0.092$~mag airmass$^{-1}$ (Nikolaev et al. 2000). We estimate a
systematic error $< 0.2$ mag for the photometric zero point. Absolute
magnitude calibration for the $K_s$-band was performed using three
2MASS stars contained in the $2\times2$ arcmin PANIC field of
view. Note that the LCO photometric system yields a $K_s$-band
magnitude that is the same within the photometric errors to that
obtained with 2MASS (see e.g. Carpenter 2001).

\subsubsection{Astrometry}

An absolute astrometry was derived by tying all the images to the
2MASS catalogue.  First, we established a plate solution for the
\fors\ $V$-band image using the {\it IRAF} tasks {\it ccmap} and {\it
cctran} on fourteen 2MASS reference stars. The $rms$ residuals of the
astrometric fit were $<0\farcs1$. Next we measured the positions of
ten foreground stars in the \fors\ image that were close to \rbs. This
secondary grid of reference stars was necessary due to the smaller
field of view covered by \isaac\ and PANIC, and because some of the
above primary references are saturated in the Blanco and Keck
frames. After accounting for all the errors of our astrometric chain
and for the intrinsic absolute astrometric accuracy of 2MASS
(Skrutskie et al. 2006), we end up with an overall uncertainty of
$<0\farcs23$ on the absolute astrometry.

\subsection{Radio observations: {\it ATNF Parkes}}
\label{parkes}

\rbs\, was observed in radio with the dual-band coaxial
10-50cm receiver of the Parkes radio telescope in 2006 April 13th .
The observations was carried out at 2.9\,GHz and 708\,MHz (10\,cm and
50\,cm, respectively) simultaneously, for a total of 3.6 hrs and data
have been one bit sampled every 0.215\,ms (Burgay et al.~2007, in
prep).

Data analysis was done with {\tt vlsa} (e.g. Burgay 2000), a code
using a standard FFT based periodicity search algorithm. The data have
been dedispersed with different trial values of the dispersion measure
(DM) ranging from 0 to 200; the range adopted is very conservative
since, assuming for \rbs\, a distance of 400\,pc (Posselt et al.~2007),
the predicted Galactic DM is $<$10\,pc\,cm$^{-3}$ (Taylor \& Cordes
1993; Cordes \& Lazio 2002).

No clear signal (having signal-to-noise ratio S/N $> 8$) was found
around the spin period detected in X--rays (Zane et al.~2005) nor at
any period ranging from 1\,ms to 10\,s. The sensitivity limits of this
search can be calculated using the radiometer equation (see Manchester
et al.~2001) with the following parameters: $T_{sys}$ equal to 30 and
40\,K , $G$ equal to 0.67 and 0.59\,K/Jy, for the 2.9\,GHz and 708\,MHz
observation respectively.

For a pulsar with period P=9.437\,s and a duty-cycle of $< 5$\% we
obtain a flux density limit $S_{min} < 0.33$\,mJy for the observations
at 708 MHz and $S_{min} < 0.06$\,mJy for the observation at 2.9\,GHz.

%%%%%%%%%%%%%%%%%%%%% FIGURE vFv %%%%%%%%%%%%%%%%%%%%%%%%%55
\begin{figure*}
\centerline{
\psfig{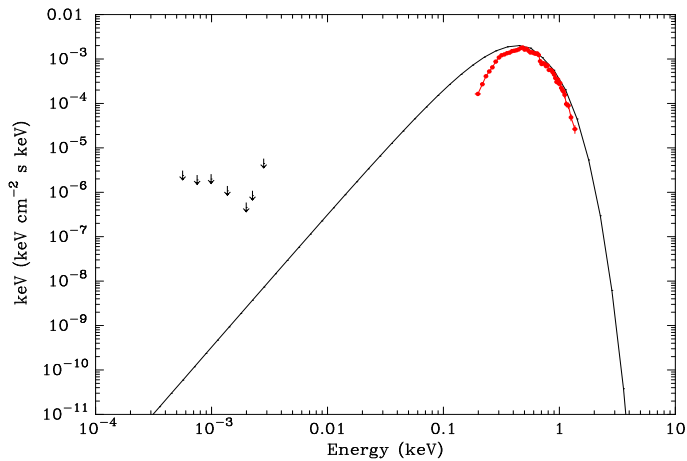}}
\caption{Multiband spectrum of \rbs . \xmm\, spectrum is plotted, labelled with filled circles (Zane et al.~2005). The solid line shows the unabsorbed blackbody which best fit the X--ray data. The arrows represent our 5$\sigma$ upper limits on the de-reddened optical and infrared flux.}
\label{vFv}
\end{figure*}
%%%%%%%%%%%%%%%%%%%%%%%%%%%%%%%%%%%%%%%%%%%%%%%%%%%%%%%%%%%%%%%%

%%%%%%%%%%%%%%%%%%%%%%%%%%%%%%%%%%%%%%%%%%%%%%%%%%%%%%%%%%%%%%%%%%%%%%

\begin{table}
\begin{minipage}{0.46\textwidth}
\begin{center}
\begin{scriptsize}
\begin{tabular}{cccccccccc}

\hline
\hline
B & V & r$^{\prime}$ & i$^{\prime}$ & J & H & K$_s$ & 2.9\,GHz & 708\,MHz \\
\hline
24.0 & 25.5 & 25.7 & 24.2 &  22.6 & 21.9 &  20.8 & 0.33 & 0.06 \\   

\hline
\hline
\end{tabular}
\end{scriptsize}
\caption{{\em Upper row}: observing bands. {\em Lower row}: Optical/infrared magnitude and radio flux $5\sigma$ upper limits of the counterpart to the XDINS \rbs. Radio flux upper limits refers to pulsed emission and are in unit of mJy  (Burgay et al. 2007, in prep).}
\end{center}
\end{minipage}

\end{table}

%%%%%%%%%%%%%%%%%%%%%%%%%%%%%%%%%%%%%%%%%%%%%%%%%%%%%%%%%%%%%%%%%%%%%%

\section{Results}
\label{results}

Thanks to the \CXO\, high spatial accuracy, we inferred a $0\farcs6$
accurate position for \rbs\, (see \S\ref{chandra}), crucial to search
for a possible optical/infrared counterpart (see also
Fig.\,\ref{iroptrbs}). Furthermore, studying the \CXO\, PSF we
confirmed with a higher accuracy the point--like nature of this
source.

We have used our improved \CXO\, position to search for the optical/infrared 
counterpart to \rbs. Unfortunately our images were not deep
enough to detect any optical or infrared counterpart (see
Fig.\,\ref{iroptrbs} and the discussion section for details). The four
objects identified close to the \cxo\ position (see
Fig.\,\ref{iroptrbs}) have been already considered unlikely candidates
by Mignani et al. (2007) on the base of their $B-V > 0$, and of their
relatively high optical brightness.  Furthermore, two of the objects
look extended and are probably extra-galactic background sources.

Note that XDINSs are characterised by high proper motions, hence
observations too far in time from our \CXO\, observation might make a
possible optical/infrared counterpart lie out of the
\CXO\, positional error circle. However, except for the Keck
observation, all of our observations were close enough to the \CXO\,
pointing not to suffer from this effect.

We report in Tab.\,2 the 5$\sigma$ upper limits on our non--detections
for each band. These limits were derived for each filter from the
magnitude of faintest stars detected at 5~$\sigma$ confidence level
among the stars that fell on the same CCD as \rbs.
We calculated the reddening in the direction of \rbs\,, from the
hydrogen absorption value inferred from the X--ray spectrum
($N_{H}=3.6\times10^{20}$\,cm$^{-2}$; Zane et al.~2005), which gives
$E(B-V)=0.062$ mag (Bohlin, Savage \& Drake~1978). We then converted
this value into an estimate of the reddening in all the filters we
actually used (Cardelli, Clayton \& Mathis~1989).

The radio observations we report here (see \S\ref{parkes}) are, up to
now, the deepest available for this source and for XDINSs in general
(Burgay et al.~2007, in prep), and the obtained radio limits are among
the most stringent to date for radio observations of XDINS (see
e.g. Kaplan et al.~2003)\footnote{Interestingly, pulsed radio emission
from \rbs\, at $\sim110$~MHz was reported by Malofeev et
al.~(2007). However, this detection still needs a confirmation. We
note that our non detection at higher frequency would imply a steep
spectrum, with spectral index $>2.1$.}.

Furthermore, in order to test the hypothesis that XDINS are related to
Rotating RAdio Transients (RRATs, McLaughlin et al. 2006), a search
for single dedispersed pulses (Cordes \& McLaughlin 2003) has also
been carried out giving negative results. However, the presence of
strong radio frequency interferences do not allow us to give a
definite answer on the presence or absence of RRAT-like emission in
this source.

\section{Discussion and conclusions}
\label{discussion}

We presented the results of a multiband observational campaign of the
field of the isolated neutron star \rbs. 
%Data were taken in several energy ranges, covering the X-ray, optical,
%infrared and radio bands. The short X--ray observation performed with
%the HRC--I camera on board of \CXO\, allow us: i) to get a
%sub-arcsecond accurate position for this source, crucial for searching
%for counterparts in other bands, and ii) to perform a detailed PSF
%study which confirmed the point--like nature of this neutron star.
No optical/infrared/radio counterparts have been found (upper
limits are reported in Tab.\,2 and Fig.\ref{vFv}) within the
very small \CXO\, error circle of the source (see
\S\,\ref{chandra}).

A feature common to all optically identified XDINSs is that their
optical flux lies well above the extrapolation of
the X-ray blackbody at lower energies (e.g. Kaplan, Kulkarni \&
van Kerkwijk~2003).  This ''optical" excess varies from source to
source and can be as high as a factor of 10.  Whether this is due to
emission from regions of the star surface at different temperatures or
to other mechanisms, such as non-thermal emission from particles in
the star magnetosphere or to a substantial suppression of the crustal
emission at X-ray energies (Zane \& Turolla~2005, P\'erez-Azor\'in et
al.~2006), is still under debate. 

Basing on the similarities with the emission of the other optically
detected XDINSs (see e.g. van Kerkwijk \& Kulkarni~2001; Pons et
al.~2002; Kaplan et al.~2003; Ho et al. 2007), \rbs\, is expected to
be quite faint. We show in Fig.\,\ref{vFv} the extrapolation of the
X-ray spectrum detected by
\xmm\, (Zane et al. 2005). Assuming an optical excess as large as
a factor 10, the expected magnitudes of
\rbs\ are B$\sim$28, V$\sim$28.3, J$\sim$29.2, H$\sim$29.3 and
K$\sim$30. Therefore, although the limits presented in this paper are
the deepest ones presently available for \rbs, they are not deep
enough to constrain its optical/infrared spectrum, and in
particular the presence of an additional cooler blackbody component
which might be responsible for the optical/infrared emission. 

It has been suggested that XDINSs can
be related to the magnetars, because of their intriguing similarities
in their spin periods and magnetic fields. As proposed in Mignani et
al.~(2007), a further piece of evidence to relate the two classes
might come from the comparison of their infrared spectra which, in the
case of the magnetars, are characterised by a distinctive flattening
with respect to the extrapolation of the X-ray spectrum (Israel et
al.~2004). If such a flattening is due to a genuine turnover in the
neutron star spectrum or due to the presence of a fossil disk, as
proposed for the magnetars, the detection of the same effect in both
classes would be extremely interesting. In this respect, \rbs\, looks
particularly promising.  By interpreting the broad spectral feature
observed at $\sim$ 0.7 keV either as a proton cyclotron resonance or
bound-bound, bound-free transitions in H or He-like atoms, it requires
a magnetar-like magnetic field, $B\sim 10^{14}$ G, the highest among
XDINSs. A systematic search for infrared emission from nearly all
XDINSs, including \rbs, did not show any evidence for a spectral
turnover redward of the $R$ band (Lo Curto et al.~2007). Our
multiwavelength campaign also shows no signs for such a
turnover. Taking as a reference the SEDs of RX J0720.4--3125 (Kaplan
et al.~2003) and RX J1856--3754 (Pons et al.~2002; Ho et al.~2007) and
using the same X-ray-to-optical normalisation (e.g. Mignani et
al. 2007), such a spectral flattening would imply $H\ge 25$ for \rbs.

%%%%%%%%%%%%%%%%%%%%%%%% radio figure %%%%%%%%%%%%%%%%%%%%%%%%%%
\begin{figure*}
\centerline{
\psfig{figure=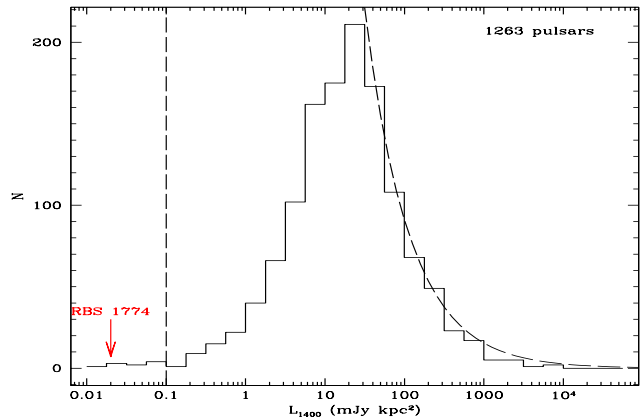,width=9cm,height=6cm}}
\caption{Distribution in luminosity at 1400 MHz of the observed pulsar
  population of the galactic field. The dashed curve is the radio
  pulsar's luminosity function (Lorimer et al. 2006). The arrow
  indicates the luminosity limit reached in the present search.}
\label{lum1400}
\end{figure*}

%%%%%%%%%%%%%%%%%%%%%%%%%%%%%%%%%%%%%%%%%%%%%%%%%%%%%%%%%%%%%

The radio observations we performed (see \S\ref{parkes}) gave us very
stringent limits, among the most stringent to date for radio
observations of XDINS. Assuming a distance of 400\,pc and a typical
spectral index for long period pulsars of 1.7 (e.g. Kramer et
al. 1998) we obtain a luminosity limit at 1400\,MHz $L_{1400}<
0.02$\,mJy\,kpc$^2$, lower than what expected for the majority of
radio pulsars, for which the most recently derived luminosity function
(Lorimer et al.~2006) has a lower limit of 0.1\,mJy\,kpc$^2$. A
comparison with the luminosity at 1.4\,GHz of the known population of
pulsars also shows that our limits are very stringent (see
Fig. \ref{lum1400}), including 99.8\% of all known non-recycled
galactic field pulsars.

From the above considerations we can hence conclude that, if \rbs\, is
active as radio pulsar, its non detection is more probably due to a
geometrical bias (i.e. the radio beam not pointing toward us) than to
a luminosity bias. If the pulse duty-cycle is, as assumed above, 5\%,
the radio beam semi-aperture is $\geq 9^\circ$, implying that the
probability of an unfavourable geometry is $\leq$ 98\%.  Note that, if
the radio beam is smaller, as suggested by the empirical law by Rankin
(1993), the geometrical bias is even worse but our flux limits become
even more stringent: for a duty-cycle of 1\%, for instance, the limits
at 2.9\,GHz and 708\,MHz are, respectively, 0.03\,mJy and 0.14\,mJy.

Considering the importance of such studies, future deeper
optical/infrared/radio observations of \rbs\ (and other XDINSs) should
be carried out to pinpoint its very faint counterparts. The knowledge
of the \CXO\, accurate position we report here, will play a crucial
role in the counterpart identification.

\section*{Acknowledgements}

We acknowledge the director of the \CXO\, X-ray Observatory, Harvey
Tananbaum, for according us the \rbs\, observation through its
Director Discretionary Time. These results are based on observations
collected at the European Southern Observatory, Paranal, Chile under
programme ID 71.C-0189(A), 075.D-0333(A), as well as data gathered
with the 6.5 meter Magellan Telescopes located at Las Campanas
Observatory, Chile. Part of the data presented here were obtained at
the W.M. Keck Observatory, which is operated as a scientific
partnership among the California Institute of Technology, the
University of California and the National Aeronautics and Space
Administration. The Observatory was made possible by the generous
financial support of the W.M. Keck Foundation. NR is supported by the
Netherlands Organisation for Scientific Research (NWO) through a
Post-doctoral Fellowship and a Short Term Visiting Fellowship of the
University of Sydney, and thanks the IoA of the University of Sydney
for the warm hospitality. MAPT is supported in part by NASA LTSA grant
NAG5-10889, PGJ acknowledges support from the NWO, SZ thanks PPARC for
financial support through an Advanced Fellowship, and DS acknowledges
a Smithsonian Astrophysical Observatory Clay Fellowship as well as
support through the NASA Guest Observer program. We thank
M.\,McLaughlin, S.\,Popov and A.\,Possenti for according us permission
to publish the Parkes data. We thank the referee, Fred Walter, for
useful suggestions.

\label{lastpage}


\begin{thebibliography}{99}

\bibitem{}
Bohlin, R.~C., Savage, B.~D., Drake J.~F. 1978, ApJ, 224, 132

\bibitem[\protect\citeauthoryear{Brazier \& Johnston 1999}{1999}]{bj99}
Brazier, K.~T.~S., Johnston, S. 1999, MNRAS, 305, 671

\bibitem[\protect\citeauthoryear{Burgay 2000}{Burgay}{2000}]{bur00}
Burgay, M., 2000, Master thesis, University of Bologna

\bibitem{}
Cardelli, J.~A., Clayton, G.~C., Mathis, J.~S. 1989, ApJ, 345, 245

\bibitem{}
Carpenter, J. 2001, AJ, 121, 2851

\bibitem{}
Geppert, U., K\"uker, M., Page, D. 2006, A\&A, 426, 267

\bibitem[\protect\citeauthoryear{Geppert, K\"uker \& Page 2006}{2006}]{urme06}
Geppert, U., K\"uker, M., Page, D. 2006, A\&A, 467, 937

\bibitem[\protect\citeauthoryear{Cordes \& McLaughlin 2003}{Cordes \& McLaughlin}{2003}]{cm03}
Cordes, J.~M. \& McLaughlin, M.~A. 2003, ApJ, 596, 1142 (2003).

\bibitem[\protect\citeauthoryear{{Cordes} \& {Lazio}}{{Cordes} \& {Lazio}}{2002}]{cl02}
Cordes J.~M. \& Lazio, T.~J.~W.,  2002, astro-ph/0207156

\bibitem{}
Cropper, M., et al. 2007, Ap\&SS, 308, 161

\bibitem{}
Damiani, F., Maggio, A., Micela, G., Sciortino, S. 1997, ApJ, 483, 350

\bibitem{}
Haberl, F. 2007, Ap\&SS, 308, 181

\bibitem{}
Ho, W., et al. 2007, MNRAS, 375, 821

\bibitem{}
Israel, G.~L., et al. 2004, IAU 218, eds. Camilo, F. \&  Gaensler, B.~M., astro-ph/0310482

\bibitem[\protect\citeauthoryear{Johnston 2003}{2003}]{john03}
Johnston, S. 2003, MNRAS, 340, L43

\bibitem[\protect\citeauthoryear{Kaplan et~al. 2003}{Kaplan et~al.}{2003}]{kkm+03}
Kaplan, D.~L., et al. 2003, ApJ, 590, 1008

\bibitem{}
Kaplan, D.~L., Kulkarni, S.~R., van Kerkwijk, M.~H. 2003, ApJ, 588, L33

\bibitem[\protect\citeauthoryear{Kaplan \& van Kerkwijk 2005}{2005}]{kvk05a}
Kaplan, D.~L., van Kerkwijk, M.~H. 2005a, ApJ, 628, L45


\bibitem[\protect\citeauthoryear{Kaplan \& van Kerkwijk 2005}{2005}]{kvk05a}
Kaplan, D.~L., van Kerkwijk, M.~H. 2005b, ApJ, 635, L65

\bibitem{}
Kaplan, D.L., van Kerkwijk, M.~H. \& Anderson, N. 2007, ApJ, in press, astro-ph/0703343

\bibitem[\protect\citeauthoryear{Kramer et~al. 1998}{Kramer et~al.}{1998}]{kxl+98}
Kramer, M., et al. 1998, ApJ,  501, 270



\bibitem{}
Lo~Curto, G., Mignani, R.~P., Perna, R., Israel, G.~L. 2007, submitted to A\&A



\bibitem[\protect\citeauthoryear{Lorimer et~al. 2006}{Lorimer et~al.}{2006}]{lfl+06}
Lorimer, D.~R., et al. 2006, MNRAS, 372, 777


\bibitem[\protect\citeauthoryear{Malofeev et al. 2005}{2005}]{malof05}
Malofeev, V.~M., et al. 2005, Astr. Rep., 49, 242

\bibitem[\protect\citeauthoryear{Malofeev et al. 2007}{2007}]{malof07}
Malofeev, V.~M., et al. 2007,  Ap\&SS, 308, 211


\bibitem[\protect\citeauthoryear{Manchester et~al. 2001}{Manchester et.~al.}{2001}]{mlc+01}
Manchester, R.~N., et al. 2001, MNRAS, 328, 17

\bibitem{}
Martini, P., et al. 2004, SPIE, 5492, 1653

\bibitem[\protect\citeauthoryear{McLaughlin et~al. 2006}{McLaughlin et~al.}{2006}]{mll+06}
McLaughlin, M.~A., et al. 2006, Nature, 439, 817


\bibitem{}
Mignani, R. P., et al. 2007, Ap\&SS, 203

\bibitem{}
Motch, C., Zavlin, V.~E., Haberl, F. 2003, A\&A, 408, 323

\bibitem{}
Motch, C., et al. 2005, A\&A, 429, 257

\bibitem{}
Neuh\"auser, R. 2001, AN, 322, 3

\bibitem{}
Nikolaev, S., et al. 2000, AJ, 120, 3340

\bibitem{}
Oke, J.~B. et al.~1995, PASP, 107, 375

\bibitem[\protect\citeauthoryear{P\`{e}rez-Azor\'in et al. 2006}{2006}]{perez06}
P\'erez-Azor\'in, J.~F., Pons, J.~A., Miralles, J.~A., Miniutti, F.
2006, A\&A, 459, 175

\bibitem{}
Persson, S.~E., et al. 1998, AJ, 116, 2475

\bibitem[\protect\citeauthoryear{Pons et al. 2002}{2002}]{pons02}
Pons, J.~A., et al. 2002, ApJ, 564, 981



\bibitem[\protect\citeauthoryear{Posselt et al. 2007}{2007}]{bettina07}
Posselt, B., et al. 2007, Ap\&SS, 308, 171


\bibitem[\protect\citeauthoryear{Rankin 1993}{Rankin}{1993}]{r93}
Rankin, J.~M. 1993, ApJ, 405, 285 

\bibitem[]{}
Skrutskie, M.~F., et al. 2006, AJ, 131, 1163

\bibitem[\protect\citeauthoryear{Taylor \& Cordes}{Taylor \& Cordes}{1993}]{tc93}
Taylor, J.~H. \& Cordes, J.~M. 1993, ApJ, 411, 674


\bibitem[\protect\citeauthoryear{Turbiner et al. 2007}{2007}]{tur07} 
Turbiner, A.~V., et al. 2007, Ap\&SS, 308, 267


\bibitem[\protect\citeauthoryear{Turbiner \& Lopez-Vieyra 2007}{2007}]{tlv07} 
Turbiner, A.~V., Lopez-Vieyra, J.~C. 2006, Phys. Reps., 
424, 309


\bibitem[\protect\citeauthoryear{Turolla et al. 2004}{2004}]{robsil04}
Turolla, R., Zane, S., Drake, J.J. 2004, ApJ, 603, 265

\bibitem{}
van Kerkwijk, M.~H. \& Kulkarni, S.~R. A\&A, 378, 986

\bibitem[\protect\citeauthoryear{van Kerkwijk \& Kaplan 2007a}{2007a}]{vkk07a} 
van Kerkwijk, M.~H., Kaplan, D.~L., 2007, Ap\&SS, 308, 191


\bibitem{}
Walter, F.~M. 2001, ApJ, 549, 433

\bibitem{}
Walter, F.~M., Lattimer, J.~M., 2002, ApJ, 576, L145

\bibitem{}
Woods, P.~M. \& Thompson, C. 2006, Compact stellar X-ray
sources. Ed. W.~Lewin \& M.~van~der~Klis, Cambridge University Press,
astro-ph/0406133


\bibitem{}
Zampieri, L., et al. 2001, A\&A, 378, L5


\bibitem{}
Zane, S., et al. 2005, ApJ, 627, 397

\bibitem{}

Zane, S. \& Turolla, R. 2005, AdSpR, 35, 1162

\bibitem{}
Zane, S., de Luca, A.,  Mignani, R.~P. \&  Turolla, R.  2006, A\&A, 457, 619


\bibitem[\protect\citeauthoryear{Zane \& Turolla 2006}{2006}]{silrob06}
Zane, S. \& Turolla, R. 2006, MNRAS, 366, 727

\bibitem{}
Zane, S. 2007, Ap\&SS, 308, 259
\bibitem{}
Zombeck, M.~V., et al. 1995, SPIE, 2518, 96

\end{thebibliography}
\end{document}